\documentclass[prl,8pt,twocolumn,nofootinbib,nobibnotes,superscriptaddress]{revtex4-1}
\usepackage{amssymb,graphicx,color,amsmath,xfrac,bm,ifthen,lineno,lipsum,caption,ulem}

\usepackage[mathletters]{ucs}
\usepackage[utf8x]{inputenc}

\definecolor{lightblue}{rgb}{0.17,0.39,1}
\definecolor{lightgreen}{rgb}{0.67,0.81,0.08}
\definecolor{lightred}{rgb}{1,0.05,0.52}

\usepackage[
bookmarks,
colorlinks=true,
breaklinks,
pdftitle={Scale-invariant magnetoresistance in a cuprate superconductor},
pdfauthor={Arkady Shekhter}
]{hyperref}

\hypersetup{
linkcolor=black,
citecolor=black,
filecolor=black,
urlcolor=black
}

\newcommand{\LSCO}{\,$\text{La}_{2-x}\text{Sr}_x\text{CuO}_4$\,}
\newcommand{\YBCO}{\,$\text{YBa}_2\text{Cu}_3\text{O}_{6+\delta}$\,}
\newcommand{\TlBCO}{\,$\text{Tl}_2\text{Ba}_2\text{Cu}\text{O}_{6+\delta}$\,}

\newcommand{\BISCCO}{\,$\text{Bi}_2\text{Sr}_2\text{CaCu}_2\text{O}_{8+\delta}$\,}
\newcommand{\BaFeAs}{\,$\text{BaFe}_2(\text{As}_x\text{P}_{1-x})_2$\,}

\renewcommand{\b}[1]{ {\bf{#1}} }

\begin{document}

\title{Scale-invariant magnetoresistance in a cuprate superconductor}

 \author{P.~Giraldo-Gallo}
\protect\thanks{These authors contributed equally}
	\affiliation{National High Magnetic Field Laboratory (NHMFL), Florida State University, Tallahassee, FL 32310, USA}
	\affiliation{ Department of Physics, Universidad de Los Andes, Bogotá 111711, Colombia} 

 \author{J.~A.~Galvis}
	\protect\thanks{These authors contributed equally}
 		\affiliation{National High Magnetic Field Laboratory (NHMFL), Florida State University, Tallahassee, FL 32310, USA}
		\affiliation{Departamento de Ciencias Naturales, Facultad de Ingenierıa y Ciencias Basicas, Universidad Central, Bogotá 110311, Colombia.}
	
 \author{Z.~Stegen}
 		\altaffiliation[Present address: ]{Northrop Grumman Corporation, Linthicum, MD 21090, USA}
 		\affiliation{National High Magnetic Field Laboratory (NHMFL), Florida State University, Tallahassee, FL 32310, USA}
 		\affiliation{Department of Physics, Florida State University, Tallahassee, FL 32310, USA}
 
 \author{K.~A.~Modic}
 		\affiliation{Max-Planck-Institute for Chemical Physics of Solids, Noethnitzer Strasse 40, D-01187,
 		Dresden, Germany}
 		
 \author{F.~F~Balakirev}
 \author{J.~B.~Betts}
 		\affiliation{Los Alamos National Laboratory, Los Alamos, NM 87545, USA}
 		
 \author{X.~Lian}
 \author{C.~Moir}
 \author{S.~C.~Riggs}
 		\affiliation{National High Magnetic Field Laboratory (NHMFL), Florida State University, Tallahassee, FL 32310, USA}
 		
 \author{J.~Wu}
 \author{A.~T.~Bollinger}
 		\affiliation{Brookhaven National Laboratory (BNL), Upton, New York 11973-5000, USA.}
 		
 \author{X.~He}
 \author{I.~Božović}
 		\affiliation{Brookhaven National Laboratory (BNL), Upton, New York 11973-5000, USA.}
 		\affiliation{Applied Physics Department, Yale University, New Haven, Connecticut 06520, USA.}
 		
 \author{B.~J.~Ramshaw}
  		\affiliation{Los Alamos National Laboratory, Los Alamos, NM 87545, USA}
 		\affiliation{Laboratory for Atomic and Solid State Physics, Cornell University, Ithaca, NY 14853, USA}
 		
\author{R.~D.~McDonald}
 		\affiliation{Los Alamos National Laboratory, Los Alamos, NM 87545, USA}
 		
 \author{G.~S.~Boebinger}
 		\affiliation{National High Magnetic Field Laboratory (NHMFL), Florida State University, Tallahassee, FL 32310, USA}
 		\affiliation{Department of Physics, Florida State University, Tallahassee, FL 32310, USA}
 		
 \author{A.~Shekhter}
		\email[correspondence to: ]{arkady.shekhter@gmail.com}
	 	\affiliation{National High Magnetic Field Laboratory (NHMFL), Florida State University, Tallahassee, FL 32310, USA}

\begin{abstract}
The anomalous metallic state in the high-temperature superconducting cuprates is masked by superconductivity near a quantum critical point. Applying high magnetic fields to suppress superconductivity has enabled detailed studies of the normal state. Yet, the direct effect of strong magnetic fields on the metallic state is poorly understood. We report the high-field magnetoresistance of thin-film \LSCO cuprate in the vicinity of the critical doping, $0.161≤p≤0.190$. We find that the metallic state exposed by suppressing superconductivity is characterized by magnetoresistance that is linear in magnetic fields up to 80 tesla. The magnitude of the linear-in-field resistivity mirrors the magnitude and doping evolution of the well-known linear-in-temperature resistivity that has been associated with quantum criticality in high-temperature superconductors.
\end{abstract}

\date{\today}
\maketitle

\setlength{\parindent}{0.5cm}
\setlength{\parskip}{0.5cm}

High-temperature superconductivity in the cuprates is born directly out of a “strange” metallic state that is characterized by linear-in-temperature resistivity up to the highest measured temperatures\cite{linear-BiSCCO, Anderson-science, Hussey-pheno, Zaanen-review}. In conventional metals, current is carried by long-lived electronic quasiparticles, which requires the scattering length not to be significantly shorter than the de Broglie wavelength \cite{IoffeRegel, Mott1972, bad-metals, Hussey2004-MIR}. In contrast, the resistivity in the strange metal state of the cuprates does not saturate or exhibit a crossover at the temperature where the inferred quasiparticle scattering length is comparable to the electronic wavelength. This behavior is sometimes referred to as “Planckian dissipation”, which suggests that the transport relaxation rate, $ℏ/τ$, (where $ℏ$ is the reduced Planck constant and $τ$ is the  relaxation time) is limited directly by the thermal energy scale $k_BT$  (where $k_B$ is the Boltzmann constant and $T$ is absolute temperature), rather than by quasi-particle interactions and lattice disorder\cite{Varma1989, Phillips2004, Mackenzie2013, Zaanen-planckian, Zaanen-review,AjiVarma,Zaanen-string, Kivelson2017,Hussey2009}. This calls into question the very existence of quasi-particles in the strange metal state. More important, it indicates scale-invariant dynamics (i.e., the lack of an intrinsic energy scale). This behavior is observed in both classes of high-Tc superconductors –- the cuprates and the pnictides \cite{Fisher2014,Hayes2016} –- but its microscopic origin and implications for superconductivity have yet to be fully understood. 

Scale-invariant transport is commonly associated with metallic quantum criticality. A characteristic energy scale is continuously tuned by an external parameter and vanishes when the tuning parameter crosses a critical value \cite{Zaanen-review}. For hole dopings below the critical point, $p<0.19$, the Hall effect in \LSCO (19) and quantum oscillations in \YBCO \cite{Taillefer-QO,Ramshaw-mass-Science2015}  provide evidence for a small carrier pocket, believed to be associated with a charge density wave\cite{CDW-Xray-Hayden-2012, CDW-Xray-Damascelli-2014, CDW-LSCO-Hayden}. By contrast, above the critical doping, $p>0.19$, quantum oscillations in \TlBCO \cite{Vignolle-over-2008} indicate a large hole-like Fermi surface, in agreement with band structure calculations \cite{OKA}. Measurements of Hall resistivity \cite{Boebinger2009, Boebinger-BLSCO-nature-2003,Taillefer-lsco-ybco-2016, Hussey-pheno}, the upper-critical magnetic field \cite{Taillefer-Hc2}  and the quasiparticle effective mass \cite{Taillefer-QO, Ramshaw-mass-Science2015, Vignolle-over-2008}, as well as the zero-temperature collapse of a line of phase transitions\cite{Bourges, Kerr, ultrasound, dichroism, secondharmonic}, suggest a quantum critical point near $p = 0.19$. At this doping the linear-in-temperature resistivity extends to the lowest temperatures \cite{Hussey2009, Boebinger1996, Zaanen-review} and therefore one might expect to access the anomalous behavior in the strange metal state in the broadest range of magnetic fields.

Magnetic fields have been instrumental in the study of both conventional and correlated metals because they couple directly to the charge carriers. Previous studies of the cuprates have made use of magnetic fields as a way of suppressing superconductivity to reveal the normal ground state properties through the magnetoresistance and quantum oscillations \cite{Taillefer-QO, Ramshaw-mass-Science2015, Vignolle-over-2008,Hussey2009, Proust2016,LSCO-logT, Boebinger1996, Boebinger-BLSCO-nature-2003, Boebinger2009, Taillefer-lsco-2016, Taillefer-lsco-ybco-2016} . The linear-in-temperature resistivity, however, suggests a strong interaction between the metallic state and the critical fluctuations associated with the quantum critical point. What has been missing is a study of how the magnetic field affects these fluctuations and thus the metallic state. To this end, we studied the electrical transport of \LSCO in high magnetic fields for a range of compositions near the critical doping, $x≈0.19$. We found a scale-invariant response to the magnetic field that is distinct from the well-understood response of charged quasi-particles to the Lorentz force in conventional metals \cite{Abrikosov,Pippard} . Strikingly, linear-in-field resistivity at high fields, together with linear-in-temperature resistivity at high temperatures, emerges as an intrinsic characteristic of the strange metal state in a cuprate superconductor. 

Figure 1 shows the in-plane resistivity ($ρ$) of a thin-film \LSCO cuprate sample at $p=0.190$ \cite{Bozovic2009, Bozovic2016,Bozovic2009,Bozovic2015,Takagi1989,Liang-doping,valueofp} in magnetic fields aligned along the crystallographic c-axis up to $80$ T. Linear-in-temperature resistivity down to the superconducting transition temperature, $T_c=38.6$ K (Figure 1E), indicates close proximity to the critical doping. Figure 1A shows that the magnetoresistance below $40$ K is linear in magnetic field over the entire normal-state field range. To quantify this observation we define the field-slope, $β(B,T) = dρ(B,T)/dB$. We observe that at $70$ T, $β(B,T)$ saturates below $25$ K (Figures 1, B and C, and Figure S3) which suggests that linear-in-field resistivity is an intrinsic property of the strange metal state. The saturation value of $β$ at low temperature and high fields in natural energy units is $β/μ_B=5.2$ $μΩ$cm/meV where $μ_B$ is the Bohr magneton. This is comparable in magnitude to the temperature-slope, $α(T) = dρ(T)/dT$, which is $11.8$ $μΩ$cm/meV in $α/k_B$ energy units. 

In conventional metals, magnetoresistance originates from the motion of electron quasi-particles around the Fermi surface under the action of the Lorentz force  \cite{Abrikosov,Pippard} . For a given Fermi surface morphology the strength of magnetoresistance is controlled by the product of the cyclotron frequency, $ω_c = eB/m^*$ (where $m^*$ is quasiparticle mass), and the quasi-particle relaxation time $τ$. Magnetoresistance generally decreases in conventional metals as τ decreases with increasing temperature. This is in contrast to what we observe in \LSCO at $p=0.190$ (Figure 1). At $80$ T, and between $4$ and $25$ kelvin we observe nearly a factor of $2$ increase in resistivity, suggesting a factor of $2$ decrease in $τ$ (Figs. 1, A and D) \cite{ybco-comment}, and yet the strength of the magnetoresistance  [$dρ(T)/dB$] at $80$ T between $4$ and $25$ kelvin is independent of temperature (Fig. 1 B and C). This indicates that at very high magnetic fields the transport relaxation rate is set directly by the magnetic field through $ℏ/τ ∝ μ_BB$. A mechanism other than the traditional picture of orbiting quasi-particles must therefore underlie the high-field magnetoresistance in \LSCO. One conclusion is that the magnetic field directly affects the dynamics of critical fluctuations that are responsible for the relaxation time \cite{Zaanen-review, Zaanen-planckian,AjiVarma,Zaanen-string,Kivelson2017} 

The smooth evolution of the temperature-slope $α(p)$ across the critical doping \cite{Hussey2009,Ando2004} is another indication of a lack of well-defined quasi-particles in the strange metal phase at high temperatures in contrast to the divergence of quasi-particle effective mass approaching the critical doping at low temperatures \cite{Ramshaw-comment}. The doping evolution of the magnitude of  $β(p)$ may provide further insight into the character of transport in the strange metal state. We measured the ab-plane resistivity in magnetic fields along the c-axis up to $55$ T in \LSCO over the range of dopings $p=0.161$ to $p=0.184$ (Figure 2). All samples in this doping range exhibit linear-in-temperature resistivity at high temperatures (Fig. 2B). The saturation value of $β(p)$ is shown in Fig. 2C along with $α(p)$ in natural energy units. Both $α(p)/k_B$ and $β(p)/μ_B$ decrease monotonically with doping in this doping range and evolve at a similar rate. The weak doping dependence of $β(p)$ and $α(p)$ approaching critical doping is in apparent contrast to the rapid increase in the Hall coefficient \cite{Taillefer-lsco-2016, Boebinger2009,Boebinger-BLSCO-nature-2003} and the divergence of the effective mass \cite{Ramshaw-mass-Science2015} as the critical doping is approached at low temperature and high magnetic fields. This again indicates that, despite the observation of quantum oscillations at low temperatures [in \YBCO \cite{Taillefer-QO,Ramshaw-mass-Science2015}  and \TlBCO \cite{Vignolle-over-2008}], the high-field, high temperature magnetoresistance in cuprates has a non--quasi-particle origin. 

It is well known that the transport relaxation rate is linear-in-temperature, $ℏ/τ ∝ k_BT$, in the fan-shaped region of the temperature-doping plane (Figure 3, magenta) emerging from the critical point \cite{Ando2004}. Our results (Figure 2) suggest that an analogous fan-shaped region exists in the magnetic field--doping plane (Figure 3, blue) where the relaxation rate is linear-in-field, $ℏ/τ∝μ_BB$. This extends a quantum critical region in field, temperature, and doping where the transport relaxation rate is set by the dominant energy scale, $ℏ/τ ∝ \max\{k_BT, μ_BB\}$, as illustrated in Figure 3 \cite{subdominant}.

These measurements establish the linear magnetoresistance at very high fields as a fundamental property of the strange metal state in the cuprates. A linear dependence on an external energy scale is not the only possible outcome of scale invariance near quantum critical point: in principle, any power-law dependence is possible. It is therefore striking that the temperature and field dependence of the resistivity in \LSCO assumes the simplest possible form. Both the cuprates and the pnictides \cite{Hayes2016}, exhibit this simple form of scale invariance, revealing another universal characteristic of high-temperature superconductors.

{\bf Acknowledgments:}  We thank J. Analytis, J.-H. Chu, N. Doiron-Leyraud, A. Finkel’stein, I. Fisher, S. Hartnoll, I. Hayes, S. Kivelson, J. Paglione, L. Taillefer, C. Varma, and J. Zaanen for discussions, and the entire 100 T operations team at the NHMFL Pulsed Field Facility for their support during the experiment. A.S. acknowledges the hospitality of the Aspen Center for Physics. The high-field resistivity measurements were performed in the 60 T long-pulse and 100 T magnet systems at the NHMFL Pulsed Field Facility, which is supported by NSF grant DMR-1157490 and the U.S. Department of Energy, Basic Energy Sciences (DOE/BES) “Science at 100 T” grant. Molecular beam epitaxy synthesis, lithography, and characterization of the samples were done at BNL, which is supported by DOE/BES, Materials Sciences and Engineering Division. X.H. was supported by the Gordon and Betty Moore Foundation’s EPiQS Initiative through grant GBMF4410. Aspen Center for Physics is supported by NSF grant PHY-1066293. 

\hypersetup{linkcolor=lightblue,citecolor=lightblue,filecolor=lightblue,urlcolor=lightblue}
\renewcommand{\doi}[1]{\href{https://doi.org/#1}{$^{\rightarrow}$}}
\renewcommand{\doi}[1]{\href{https://doi.org/#1}{{doi:#1}}}

\newcommand{\arxiv}[1]{\href{https://arxiv.org/abs/#1}{{arxiv:#1}}}
\renewcommand{\title}[1]{}
\newcommand{\btitle}[1]{#1}
\renewcommand{\emph}[1]{{\it{#1}}}

\cleardoublepage

\begin{widetext}


\begin{figure}[ht!!!!!!!!]
\centerline{\includegraphics[width=0.9\textwidth]{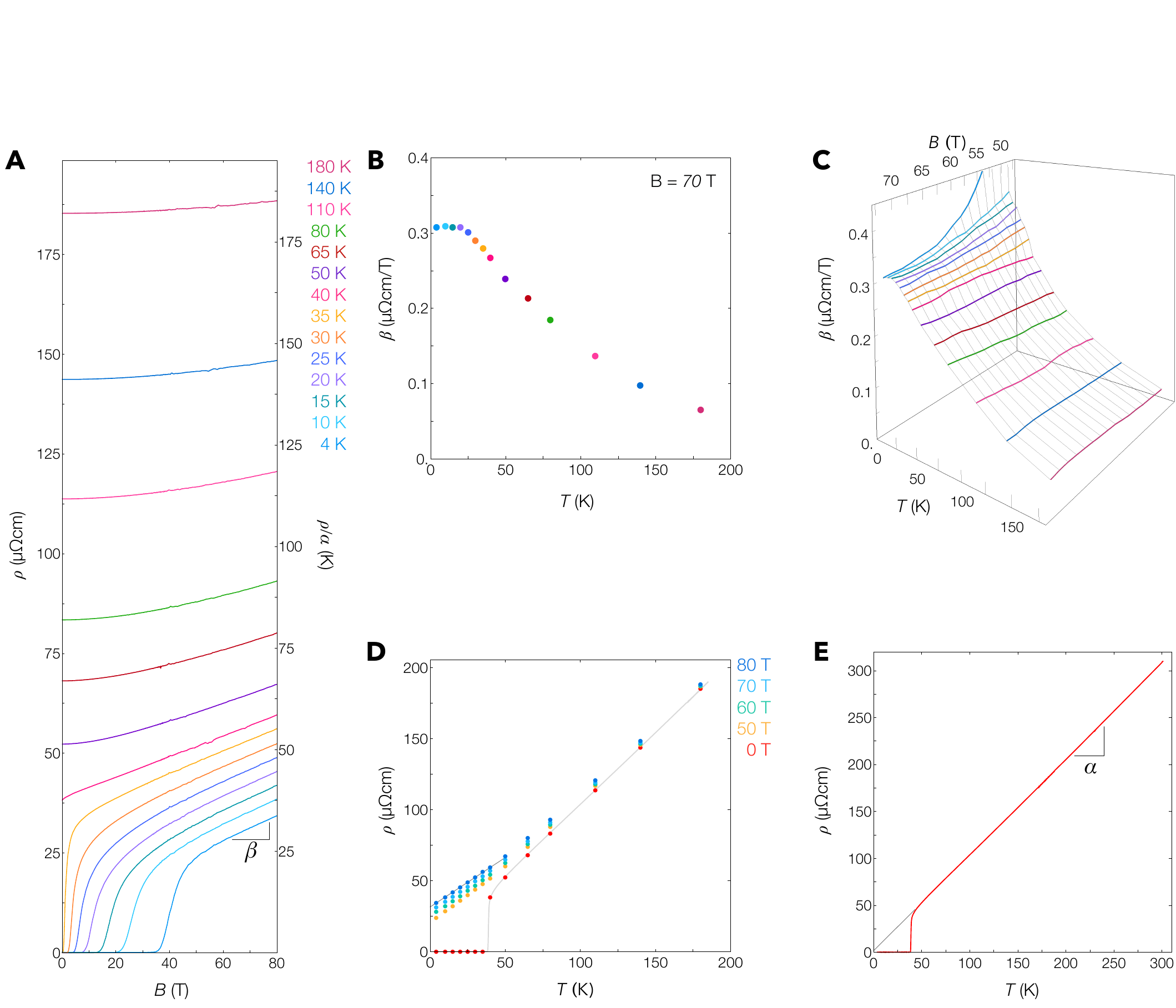}}
\caption*{ 
\b{Figure 1. ab-plane resistivity of the thin-film \LSCO at $p=0.190$. } 
The magnetic field is applied along the c-axis. 
\b{(A)} Magnetoresistance up to $80$ T for temperatures ranging from $4$ K up to $180$ K. The right axis indicates the resistivity in temperature units, $ρ/α$, where $α$ is obtained from the linear fit in (\b{E}). The aspect ratio reflects natural energy units for the magnetic field, $μ_BB$,  and temperature, $k_BT$, where the energy of $80$ T corresponds approximately to that of $53.7$ K. 
(\b{B}) Temperature dependence of $β(B,T) = dρ/dB$ at a fixed field of $70$ T obtained as slope of a linear fit for magnetoresistance in panel (\b{A}) in the field range between $65$ and $77$ T. $β(B,T)$ saturates below about $25$ K. Color-coding for temperature values as indicated in (\b{A}) also applies to (\b{B}) and (\b{C}). 
(\b{C}) Magnetic field dependence of $β(B,T)$, showing that $β(B,T)$ saturates for $B >50$ T in a broad temperature range, $10$ K $< T  < 25$ K. 
(\b{D}) Temperature dependence of the resistivity at fixed fields. The gray line indicates the zero-field resistivity from (\b{E}). 
(\b{E}) Zero-field resistivity up to room temperature. The gray line indicates a linear-fit extrapolation of the resistivity to temperatures below the superconducting transition, $ρ=ρ_0 +αT$. The magnitude of the intercept $ρ_0 ≈1.5 (\pm2)$ $μΩ$cm, and the temperature-slope, $α≈1.02 (±0.01)$ $μΩ$cm/K, are found from a linear fit in a broad temperature range. 
}
\label{fig:1}
\end{figure}

\cleardoublepage
\begin{figure}[ht!!!!!!!!]
\centerline{\includegraphics[width=0.9\textwidth]{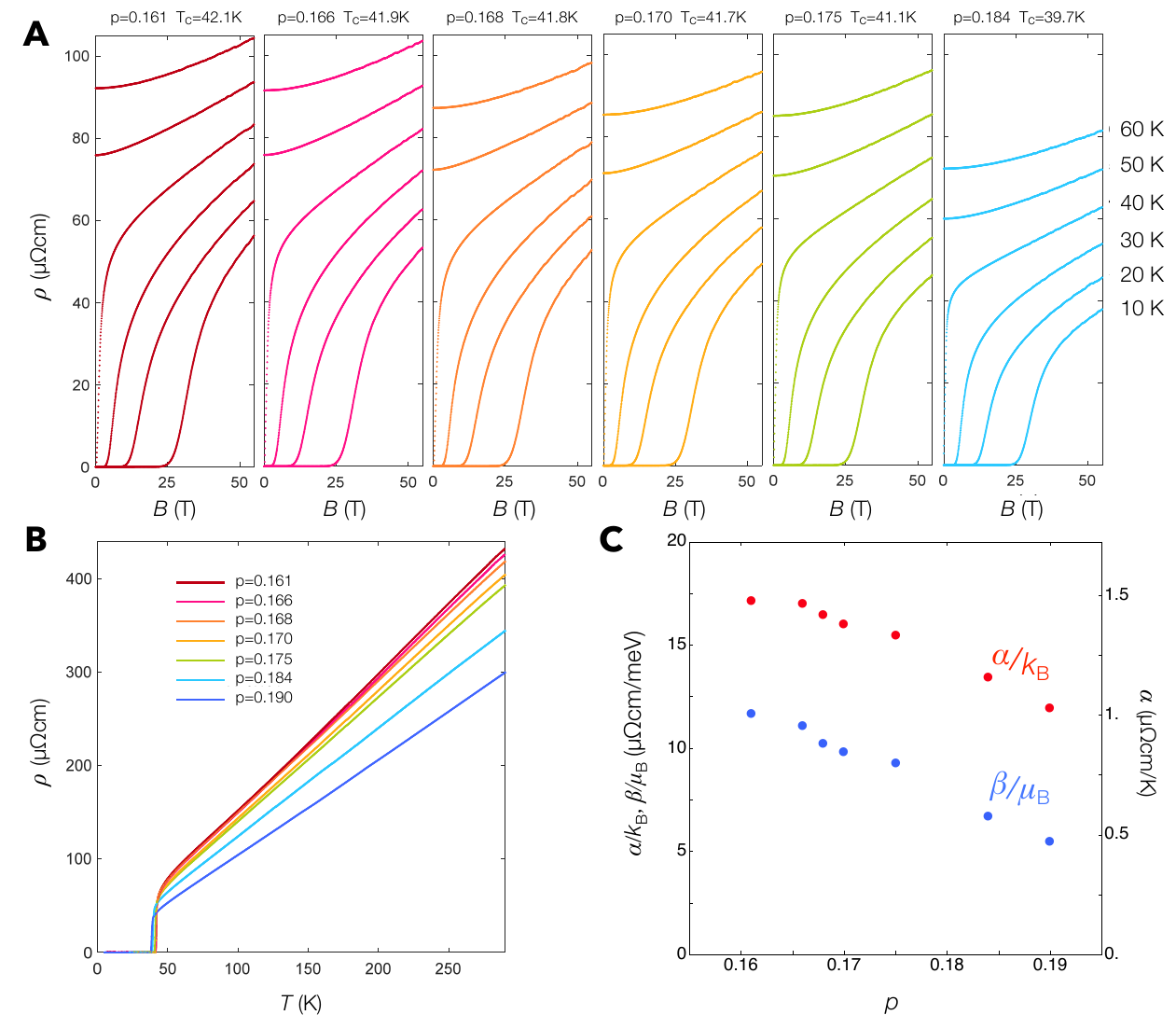}}
\caption*{
\b{Figure 2. Doping dependence of the ab-plane magnetoresistance in \LSCO.}
\b{(A)} Magnetoresistance for magnetic fields applied along the c-axis up to $55$ T in the range of dopings $0.161 < p < 0.184$ from $10$ K to $60$ K. 
\b{(B)} Zero-field resistivity versus temperature for the same set of dopings as well as doping at $p=0.190$. 
\b{(C)} Doping evolution of the temperature-slope $α(p)$  (red) and field-slope $β(p)$ (blue) in the doping range $p=0.161$ to $0.190$. The left axis indicates values of $α/k_B$ and $β/μ_B$ in energy units ($μΩ$cm/meV). The right axis indicates the value of $α$ in $μΩ$cm/K. 
}
\label{fig:2}
\end{figure}

\cleardoublepage
\begin{figure}[ht!!!!!!!!]
\centerline{\includegraphics[width=0.6\textwidth]{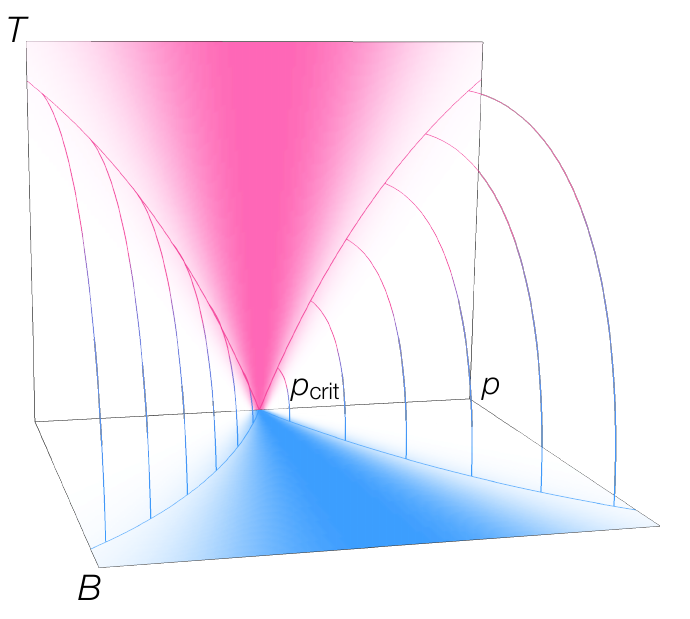}}
\caption*{
\b{Fig. 3. Schematic doping-field-temperature $(p-B-T)$ phase diagram in the vicinity of the critical doping $p_{crit}$.} Note that the superconducting phase surrounding the critical point is not shown. The magenta lines indicate the extent of the fan-shaped region (shaded in magenta) in the $p-T$ plane where linear-T resistivity exists. The fan-shaped region of linear-B resistivity in the $p-B$ plane (shaded in blue) is bounded by the blue lines. The gradient-colored lines separate the region of the $p-B-T$ space where scale-invariant transport behavior, $ℏ/τ ∝ \max\{k_BT, μ_BB\}$, exists. In the region behind these lines, a large intrinsic energy scale suppresses the anomalous dependence of $ℏ/τ$ on temperature and magnetic field. All line in this drawing indicate a smooth crossover region, not a distinct phase boundary.  
}
\label{fig:3}
\end{figure}

\cleardoublepage

\appendix

\section{ Supplementary Materials }

\subsection{ Sample Preparation}

High-quality single-crystal thin film \LSCO were grown by the combinatorial molecular beam epitaxy (COMBE) technique \cite{Bozovic2015}. The substrates are $1$ cm $× 1$ cm $× 1$ mm LaSrAlO$_4$ single crystals cut perpendicular to the crystallographic [001] direction. A copper-oxide superconductor film (of $25$ nm thickness for the highest $2$ dopings and $16.5$ nm for the rest) is patterned into a high-aspect-ratio ($300$ $µ$m × $9.8$ mm) strip with a strontium gradient along the longer edge of the strip. The Sr content changes by 4\% from one end to the other. Current leads are attached to the opposite ends of the strip. A total of $62$ voltage leads, separated from one another by $300 µ$m, are patterned on the sides of the strip, allowing longitudinal voltage measurements on segments along the strip with different strontium content. 

The high magnetic field measurements were done at the Pulsed Field Facility of the National High Magnetic Field Laboratory, at Los Alamos National Laboratory. Two different magnet systems were used for our measurements. The large gradient samples were measured in the $60$ T Controlled Magnet Waveform system with a $32$ mm size bore, that achieved $55$ T tesla pulses of several seconds duration during our magnet run. We have then chosen a sample closest to the critical doping ($x≈0.19$) for the measurements in the $100$ T Multi-shot Magnet system. This magnet system has a smaller bore ($10$ mm) and shorter pulse (few microseconds), therefore a smaller segment of the sample (about $3$ mm $× 3$ mm) was cut out for measurements. 

Figure S1A shows the superconducting transition temperatures for nominal strontium compositions, $x$, in the range from $x≈0.16$ to $x≈0.19$. Here, $T_c$ is defined as the maximum in the temperature-derivative of resistivity (Figure S1B). For the compositions studied in this work, Tc tracks closely the empirical formula $T_c/T_{c,max} = 1 − b(x − 0.16)^2$ \cite{Takagi1989,Liang-doping} , where $T_{c,max} = 42.1$ K is the maximum critical temperature for thin-film \LSCO and $b = 100$ is obtained from best fit to the data (shown in red in Figure S1 A). The value of hole-doping p for each sample is obtained from Tc using best fit formula as described above. The actual difference between nominal value of x (nominal strontium content) and the inferred value of hole doping $p$ is in fact quite small small, less than 0.005 (Figure S1A). 

The half-width at half-maximum of the temperature derivative of the resistivity curve is around 1 K for all samples studied in this work (Figure S1B for two samples). All samples show small residual resistance (less than $10$ $µΩ$cm - Figure S1c) as determined by extrapolation of the linear-in-temperature fit in the range between $150$ K and $300$ K.

Figure S2 shows the resistivity plotted as a function of $B^2$ for multiple compositions and temperatures, a superset of the data presented in Figure 2 of the main text. In the $p=0.190$ sample, a $B^2$-magnetoresistance over the entire field range ($80$ T) is observed only at the highest measured temperatures ($140$ K and $180$ K). All lower temperatures the magnetoresistance shows clear negative curvature, a direct consequence of a slower-than-$B^2$ increase at high magnetic fields.

\subsection{Running-window slope analysis}

For the $p=0.190$ sample where we have measured magnetoresistance up to $80$ T, the field-slope $β$ saturates at low temperatures well before the field limit is reached (Figure 1b,d and Figure S3). For all other samples the high-field magnetoresistance was only measured up to $55$T -- not high enough for $β$ to reach saturation at low temperatures. The comprehensive magnetoresistance dataset at multiple temperatures in these samples is redundant enough for reliable estimate of the saturation value of the field-slope $β(p)$. The simplest method using no modeling at all takes for $β$ the value of $dρ/dB$ at the highest field for temperatures in the interval between $20$K and $30$K ($β_4$ in Figure S4). In this temperature interval the magnetoresistance is linear-in-field in the broadest field range in all measured samples: superconductivity becomes important at lower temperatures while the competition between magnetic field and temperature kicks in at higher temperatures. Fig. 2C of the main text uses the values of $β$ estimated using this simple method. 

Estimates of saturation values of $β$ (for samples measured up to $55$ T) using a more quantitative analysis supports this qualitative estimate. Figure S3 illustrates three different quantitative estimates. $β_1(T)$ is obtained through the value of $dρ/dB$ at $80$ T (black circles), which for samples other than $p=0.19$ can be well approximated by extending the measured data using an empirical formula $dρ/dB = c_0 + c_1e^{−c_2B}$. $β_2(T)$ is obtained through the value of $dρ/dB$ at $55$ T (red diamonds). $β_3(T)$ is only calculated for the $p=0.19$ sample, via a linear fit for $dρ/dB$ in the field interval $65$ T and $77$ T, for which this quantity is either field-independent or depends weakly on field for all measured temperatures (blue squares). The zero-temperature limit of all βi=1..4(T) can be estimated by fitting each $β_{i=1..4}(T)$ to an Interpolating formula $β_i(T) = b_0 + b_1T/ \sinh(b_2T)$, which captures correctly the behavior at both limits and satisfactorily interpolates the behavior of  $β_{i=1..4}(p)$ in the $p=0.19$ sample in the entire field range up to $80$ T. Figure S4 shows the values of $β_{i=1..4}(p)T⟶0$ plotted vs hole-doping $p$.

\subsection{ Temperature Dependence of Resistivity at High-Field}

Figure S5 shows resistivity as a function of temperature for all compositions studied in this work (the plots are identical to Figure 2A of the main text except here each composition is shown on a separate panel). Resistivity as a function of temperature for $p=0.190$ (Figure 1 E of the main text) is not shown here. The red curve shows the zero-field resistivity. The resistivity curves are in good agreement with a previous study in bulk crystals by Cooper et al \cite{Hussey2009} . Superposed on each plot is $ρ(T)$ at $55$ T (blue circles) which shows monotonic decrease in resistivity with decreasing temperature for all studied samples \cite{Taillefer-lsco-2016}.

\cleardoublepage
\begin{figure}[ht!!!!!!!!]
\centerline{\includegraphics[width=0.9\textwidth]{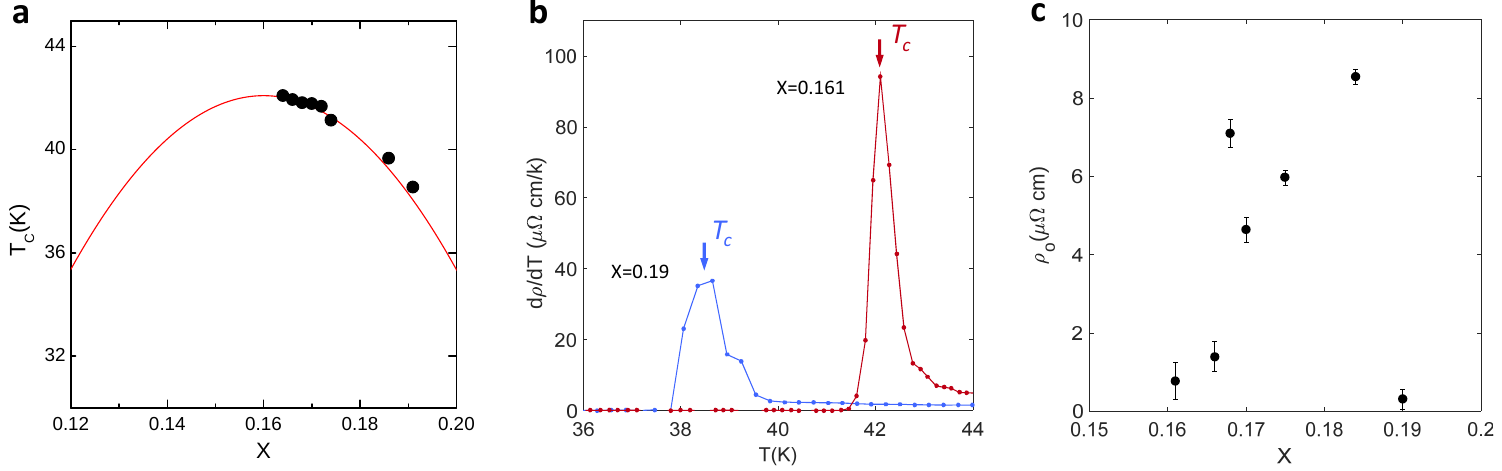}}
\caption*{
\b{Fig. S1. Characterization of the hole-doping p and residual resistivity in all samples studied in this work.} 
(A) Transition temperature $T_c$ for all samples in this work plotted vs nominal strontium content $x$ (black points). Tc is determined as the midpoint transition (maximum $dρ(T)/dT$ \cite{Liang-doping}). The transition temperatures vs strontium content track closely the empirical fit \cite{Takagi1989,Liang-doping} $T_c/T_{c,max} = 1 − b(p − 0.16)^2$ (red curve), where $T_{c,max} = 42.1$ K is the maximum transition temperature in thin-film \LSCO and $b = 100$ is the best fit parameter.
(B) Transition width of about $1$ K for two representative samples. (C) Residual resistivity, ρ0, plotted vs p for all the samples studied in this work. $ρ_0$ is determined through the extrapolation of the linear-fit to resistivity in the $150$ to $300$ kelvin range. 
}
\label{fig:S1}
\end{figure}

\cleardoublepage
\begin{figure}[ht!!!!!!!!]
\centerline{\includegraphics[width=0.9\textwidth]{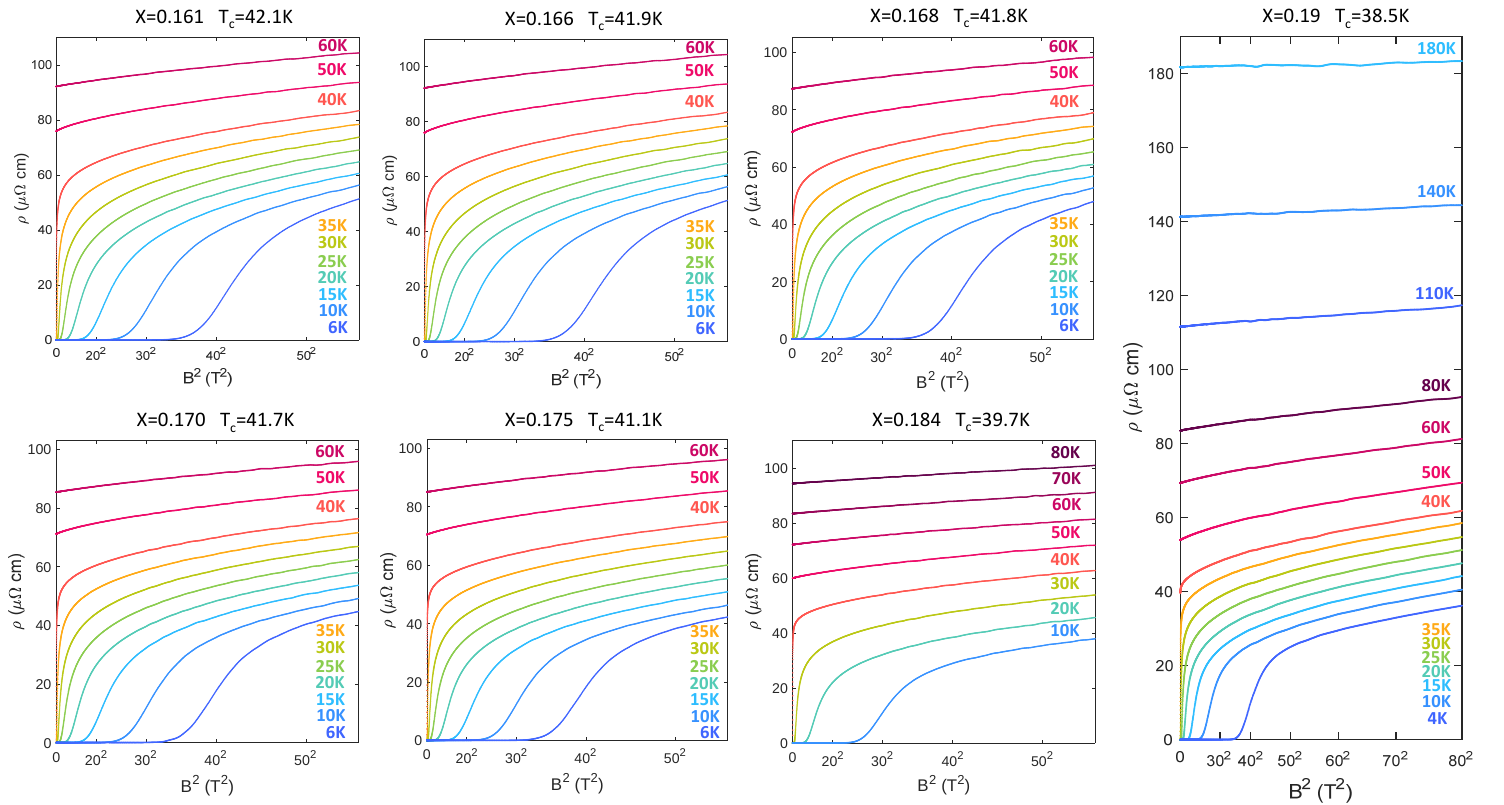}}
\caption*{
\b{Fig. S2 Magnetoresistance plotted vs B2 for all samples studied in this work.} 
Color-coding of temperature is indicated in the legend in each panel. The field range is $80$ T for $p=0.190$ sample, and $55$ T for the rest. All samples show sub-bilinear field dependence at highest measured fields. 
}
\label{fig:S2}
\end{figure}

\begin{figure}[ht!!!!!!!!]
\centerline{\includegraphics[width=0.9\textwidth]{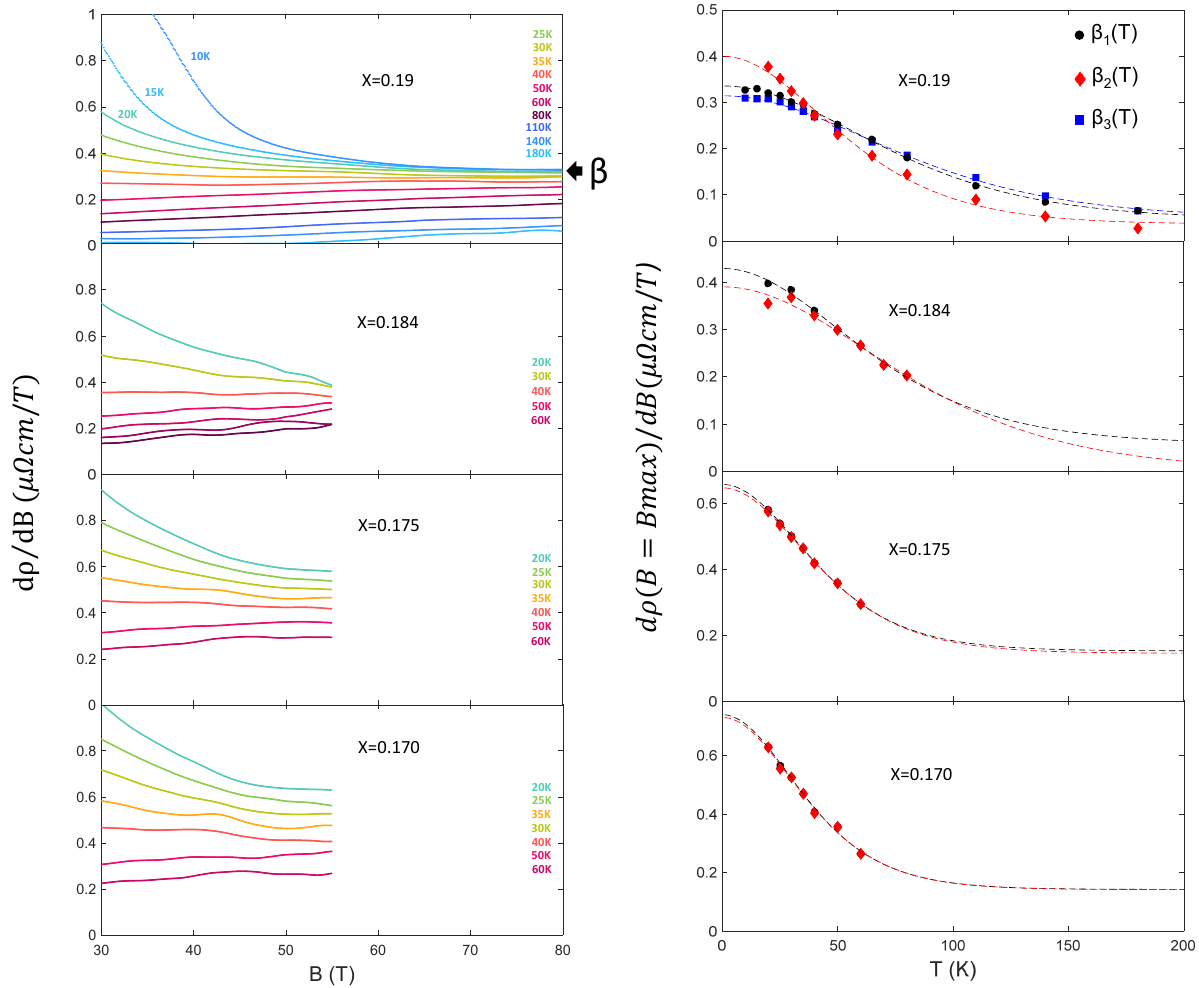}}
\caption*{
\b{Fig. S3 Estimating the saturation value of β for samples measured up to 55 T.} 
Panels on the left show $β(T,B)$ vs magnetic field B for the four highest-doping samples in this work. Temperature is color-coded as indicated in the legend in each panel. For each sample on the left, the panel on the right shows the high-field value of $β_i(T)$ plotted vs temperature (each $β_i(T)$ is obtained in different method as described above). $β_1(T)$ (black circles) is the value of $β(T,B)$ at $80$ T. $β_2(T)$ (red diamonds) is the value of $β(T, B)$ at $55$ T. $β_3(T)$ (blue squares) is a slope of a linear fit of $ρ(B,T)$ in the field interval $65$ T to $77$ T. The dashed lines represent fits for $β_{i=1..3}(T)$ with the interpolating expression $β_i(T) = b_0 + b_1T/\sinh(b_2T)$.
}
\label{fig:S3}
\end{figure}

\cleardoublepage
\begin{figure}[ht!!!!!!!!]
\centerline{\includegraphics[width=0.5\textwidth]{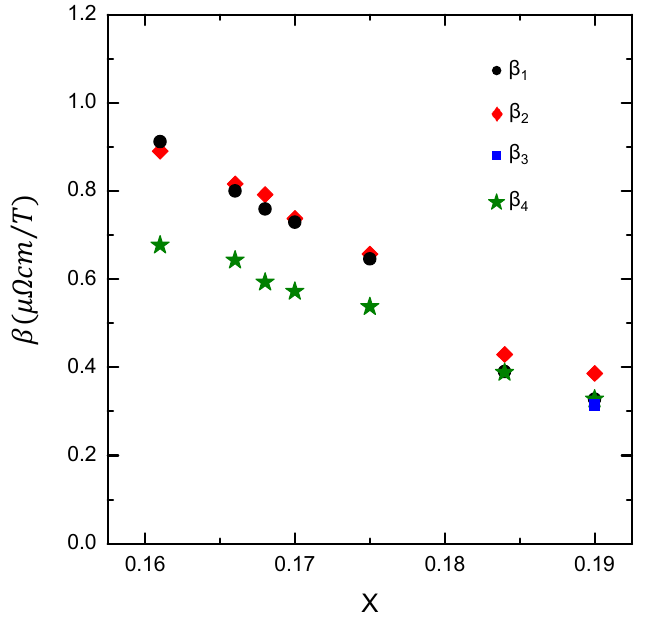}}
\caption*{
\b{Fig. S4. Comparison of different ways of estimating the saturation value of $β$}. 
See discussion in the text of SM. 
}
\label{fig:S4}
\end{figure}

\cleardoublepage
\begin{figure}[ht!!!!!!!!]
\centerline{\includegraphics[width=0.8\textwidth]{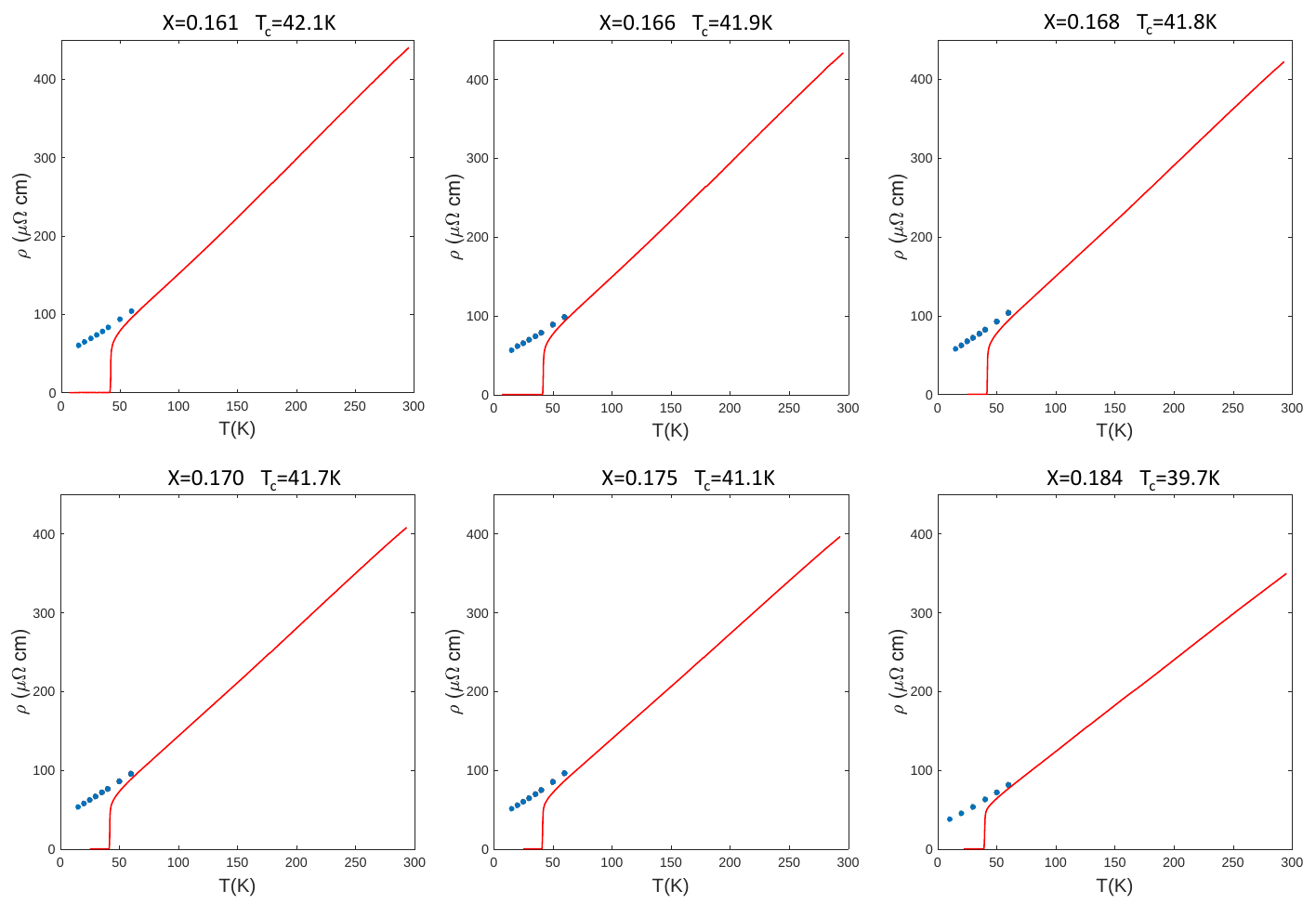}}
\caption*{
\b{Fig. S5 Temperature dependence of resistivity for all lower doping samples studied in this work.}
 Zero-field resistivity is shown in red. Blue points represent the resistivity at $55$ T. In all samples the high-field resistivity monotonically decreases with decreasing temperature
}
\label{fig:S5}
\end{figure}

\end{widetext}


\begin{thebibliography}{99}
\bibitem{valueofp} The value of p throughout this manuscript is determined from superconducting transition temperature in a broad doping range (Figure S1).
\end{thebibliography}
\end{document}